\documentclass[submission,copyright,creativecommons]{eptcs}
\providecommand{\event}{UITP2014} 

\usepackage[utf8]{inputenc}
\usepackage{amsfonts}
\usepackage{amsmath}
\usepackage{graphicx}
\usepackage[usenames,dvipsnames]{xcolor}
\usepackage{multirow}
\usepackage{wrapfig} 
\usepackage{xspace}
\usepackage{soul} 

\newcommand{\pt}{\textsc{ProofTool}\xspace}

\newcommand{\TODO}[1]{\textcolor{red}{ TODO: #1 }}

\newcommand{\comment}[1]{}

\title{Advanced Proof Viewing in \pt}
\author{Tomer Libal
\institute{Microsoft Research -- Inria Joint Center, \\ \'{E}cole Polytechnique.}
\email{shaolin@logic.at}
\and
Martin Riener\thanks{Supported by the Vienna PhD School of Informatics.}
\institute{Institute of Computer Languages, \\ Vienna University of Technology.}
\email{riener@logic.at}
\and
Mikheil Rukhaia\thanks{Supported by the project No. 52/01 of the Shota Rustaveli National Science Fundation.}
\institute{Institute of Applied Mathematics, \\ Tbilisi State University.}
\email{mrukhaia@logic.at}
}
\def\titlerunning{Advanced Proof Viewing in \pt}
\def\authorrunning{T.Libal, M.Riener, M.Rukhaia}

\begin{document}
\maketitle

\begin{abstract}
Sequent calculus is widely used for formalizing proofs.
However, due to the proliferation of data,
understanding the proofs of even simple mathematical arguments soon becomes impossible.
Graphical user interfaces help in this matter, but since they normally utilize Gentzen's original notation, some of the problems persist.
In this paper, we introduce a number of criteria for proof visualization which we have found out to be crucial for analyzing proofs.
We then evaluate recent developments in tree visualization with regard to these criteria
and propose the Sunburst Tree layout as a complement to the traditional tree structure.
This layout constructs inferences as concentric circle arcs around the root inference, allowing the user to focus on the proof's structural content.
Finally, we describe its integration into \pt and explain how it interacts with the Gentzen layout.

\comment{  Formal mathematical proofs, even of simple arguments, are often huge.
  Gentzen's sequent calculus is still widely used, even though already a dozen of inferences fail to fit onto a sheet of paper.
  Graphical user interfaces like \pt also suffer from similar problems.
  This paper tries to catch up with recent developments in tree visualization and proposes the Sunburst Tree layout as a complement to the traditional tree structure.
  It constructs inferences as concentric circle arcs around the root inference, thereby achieving a compact global view.
  We further describe the integration into \pt and the interaction with the Gentzen layout.}
\end{abstract}

\section{Introduction}

The need for visualizing data precedes the invention of computers.
Even so, the large data processed by computers made this need more explicit and initiated much research in data visualization and particularly in tree visualization.
For example, the traditional disk usage analyzers were all implemented as trees, with directories and files represented by nodes and
edges denoting the containment relation. In the last decades, the increase in disk space and the increase in number of files that followed,
prompted the design of new tree visualization methods which will be more space efficient.
One of the first methods was TreeMap~\cite{Shneiderman91} which divides a box into several smaller boxes representing the subtrees. Other algorithms made the nodes implicit by drawing fractals~\cite{Meier1996, Rosindell2012}, added a third dimension~\cite{munzner1997,huang2007,Linsen2011} or used hyperbolic and other radial approaches to better group subtrees~\cite{Lamping1995,Yang2002, Hida2005,  Stasko2000a}. Treevis.net~\cite{schulz2011}, a visual bibliography of tree viewers, now contains more than 270 different algorithms.

GAPT\footnote{General Architecture for Proof Theory, \url{http://www.logic.at/gapt}} is a framework providing data-structures,
algorithms and user interfaces for analyzing and transforming formal proofs.
The framework is very general and implements the basic data structures for simply-typed lambda calculus, for sequent and resolution proofs as well as expansion proofs.
Various theorem provers have already been integrated into this framework~\cite{PxTP}.
In parallel, we have developed a Graphical User Interface called \pt~\cite{UITP} which can be used both
as a pure visualization tool (with the features like zooming, scrolling, searching, etc.)
and as a proof manipulator (allowing to call GAPT's proof transformations such as cut-elimination, regularization, skolemization, etc.).
We are continuously extending and improving the system and one such extension was presented in~\cite{HLRR13}.

Sequent calculus proofs are often depicted as trees and in fact, the tree representation was used from the very beginning.
Gentzen's representation for sequent calculus proofs can be seen as a variant of an algorithm by Donald Knuth~\cite{knuth1971}.
The child nodes are horizontally aligned in the distance of the width of their
respective subtrees with their parent node being aligned centrally between them. The vertical alignment is determined by the distance from the root.

However, although this presentation seems natural, it is not well suited for large proof
as their structure is no longer visible. Even tracing the ancestors of a formula is cumbersome,
since the distance between parent inferences can be very large.

Despite the abundant research done in the field of tree visualization and the fact that proofs are normally represented as trees, little was
done so far in integrating these advancements into tools for proof visualization.
In fact, the first viewer which was integrated into \pt\ was a traditional tree viewer.
Proof General~\cite{aspinall07}, which also manages the proof visualization for provers such as Coq~\cite{huet97}, supports the traditional tree view as well. Other systems like $L\Omega UI$~\cite{W1} and Theorema~\cite{RISC4536} provide a structural overview in form of a DAG and a tree, respectively. However, their main focus lies on human-readable proofs where the formula level is directly contained in the text. One of the few graphical user interfaces which deviates is IDV~\cite{Trac2007109}. It renders DAG proofs in the TSTP format~\cite{SS98} using the spring layout~\cite{battista1994algorithms}.  This layout turned out to be insufficient for our needs as is discussed in~\cite{UITP}.
The reason that many advancements in tree visualization are only slowly reaching the proof theory community may primarily lie with the fact that only few of the provers care about a visual presentation of the generated object,
if they generate it at all.
Nevertheless, proof visualization is a crucial tool for analyzing large proofs like the ones we encounter in our work.
Therefore, we find it important to search for and integrate efficient tree viewers.

In this paper we propose some criteria for visualizing sequent calculus proofs and use them to analyze the existing layouts.
We argue that Sunburst Trees~\cite{Stasko2000a} are the most adequate layout and develop a new viewer for \pt,
the graphical user interface of the GAPT framework, which is based on them.
The viewer allows the displaying of the structure of the whole proof at once, to easily identify similar subproofs,
to zoom in to relevant parts and to see the relevant inference details.
At the same time it is connected to the classical Gentzen layout, which allows
 the user to focus on a small number of inferences or to be able to see an aspect of the proof which is better displayed using
the traditional view.
We believe that the visualization of proofs using a structural viewer will be useful for tools other than \pt.
In this paper we therefore  present  the benefits of using such a viewer and demonstrate its integration within \pt .

The paper is organized as follows: Section~\ref{sec:criteria} introduces the requirements we find most important in proof visualization.
In Section~\ref{sec:choosing} we explain our choice of the Sunburst Tree.
Section~\ref{sec:pt} is devoted to the integration of our Sunburst viewer into \pt and to a comparison between the two views.
Section~\ref{sec:imp} gives a description of the implementation details and
we conclude the paper in Section~\ref{sec:conc}.

\section{Criteria for Visualizing Sequent Calculus Proofs} \label{sec:criteria}

When using the traditional tree layouts, wide node labels often stretch the width of the tree and deform its structure.
One reason for that is that the context formulas of an inference need to be repeated along many branches.
Moreover, it can also happen that the main or auxiliary formulas become overly wide themselves.
Therefore, it is helpful to completely separate the tree structure from the information about the inference itself.
Since sequent calculus and the inference viewer work well on the sequent level, we concentrate on the requirements for the structural layout.

In this section, we compile a small set of requirements which were identified as critical for our proof  analysis.
We have tried comparing our conclusions with other works concerning the aesthetics of mathematical proofs.
Surprisingly, they often stay only on an abstract level. Hardy~\cite{hardy1992mathematician} names \textit{unexpectedness},
\textit{inevitability} and \textit{economy} as aesthetic properties.
The first refers to an element of surprise when a conclusion is reached, which has similarities to narratives~\cite{cain2010}.
The second stands for a detailed, convincing deduction whereas the third means restricting a proof to the minimal steps in order to prove the theorem.
Some works deepen these concepts and introduce case studies~\cite{paterson2013},
but we know of no work which details the relation between graphical notations and the aesthetics of a proof.
Therefore, the requirements given here are the result of the authors' own involvement with the formalization of mathematical proofs.

\begin{enumerate}
  \item \emph{Displaying of large proofs} is one of the most important factors. Proofs containing thousands of inference steps can become very hard to read. We would
like to find the most efficient tree representation. For example, despite the fact that proofs are traditionally denoted as trees,
the edges between the nodes play a very small role and are not space efficient.
Another aim is to be able to represent the full proof on a single screen in a comprehensive way.
This is not only useful for exporting purposes, but for tracking changes after the application of proof transformations, such as substitution or
skolemization.
  \item \emph{Distinguishing between different kinds of rules} is important as some rules, like instantiation rules, give information about the content of the proof while others,
    like contractions, give information about the shapes of proofs. Different coloring of rules is one way to distinguish between them.
  \item Without \emph{easy navigation}, one would not be able to follow the logical progress of the proof. The sub-proof relation should always
    be obvious and easy to navigate.
  \item In many cases, \emph{formula ancestor information} is important in order to relate a sequent with the atomic formulas by which it is implied.
  \item Proofs have many uses and one would sometimes like to \emph{focus on different aspects of the proof}. Proof complexity, different instantiations, cuts complexity and
    contractions may all be important for a prospective viewer of a proof. 
  \item \label{req:shape} The ability to relate \emph{shape of proofs} and sub-proofs to their content might also be an important factor since it might allow us to detect redundancies and similarities of content.
\end{enumerate}

\section{Choosing the proper tree visualization}
\label{sec:choosing}

One of the most comprehensive bibliographies for research on tree visualization is Treevis.net~\cite{schulz2011} which contains over 270 different algorithms.
Consequently, it is a challenge to pick an adequate algorithm out of the numerous ones which have been published.
However, Treevis.net also provides a categorization of the techniques in terms of the criteria of dimensionality (2D, 3D or hybrid),
representation (implicit, explicit or hybrid) and alignment (axis-parallel, radial or free). In this section we will first explain our choices with regard
to the categories mentioned. We then identify the algorithm satisfying our category requirements and show that it also meets our visualization criteria.

We decided to focus on two-dimensional representations, since there is no general additional structure which could be mapped to the third dimension.
Although both explicit and implicit representations of edges would suit our purposes, the later allows us to expect a more compact layout.
Consequently, we would like to focus on this case.
This is additionally motivated by the fact that sequent calculus proofs often contain a high amount of unary rules, where the edge is then redundant.
From the algorithms meeting these requirements, we now excluded those which do not meet criterion~\ref{req:shape} from Section \ref{sec:criteria}.
The remaining options were unexpectedly low in number.
A reason for that is that the large class of layouts based on TreeMap divides a box into equal sub-parts for each subtree.
The problem there is that in a series of identical subproofs connected by binary inferences, each subproof has half the size of the preceding one, making it nigh impossible to recognize their similarity. Fractal layouts have similar problems, whereas grid embeddings~\cite{Youn1988, Rusu2007a} do not reflect the similarity of subtrees.

What remained were explicit axis-parallel layouts and implicit radial layouts.
The first class consists of improvements on the classical Tidy Tree algorithm~\cite{Wetherell1979},
whereas the later centers around representing the tree from the root outwards.

Of special appeal to our applications was the Sunburst Tree~\cite{Stasko2000a}.
It is particularly efficient for \emph{displaying large proofs} due to its radial shape and the fact that it eliminates all edges.
One can easily \emph{distinguish different kinds of rules} by setting different colorings.
The user can group the rules by their function in the proof and thus separate the proof into parts with logical, equational or quantifier inferences.
Together with the branching structure, (sub-)proofs are already distinguishable from each other without referring to the formula level.
\emph{Navigation} is similarly simple. A single click into a subproof shrinks the original proof to half its size. At the same time the selected subproof is projected onto the circle around the full proof, giving it sufficient space to see detailed inferences. 
Sometimes, this kind of stretching impedes the identification of similar structures within the proof. In these cases, it is also possible to right-click on a node and select the ``Focus on subproof`` menu item to open a new window using the selected inference as root node.
 In order to \emph{focus on different aspects of  proofs}, a customization of two parameters is possible.
By changing both the coloring scheme as well as the inference width ratio, one can single out instantiations denoted by weak quantifiers,
different subproof complexities and other aspects. Another strong point of the Sunburst viewer is its ability to relate content
to \emph{proof shape}. This is again achieved by inference coloring and width ratio and by its efficient presentation of a whole proof on a single screen.
Finally, a radial layout is also helpful in that it can always be drawn into a square, leaving room on the screen for the inference information.

There is one requirement where the Sunburst viewer falls far behind the traditional tree viewers and this is with keeping \emph{formula ancestor information}.
The relationship between a formula and its ancestor, while easily displayed in the traditional viewer, cannot be represented in Sunburst.
This raises another important requirement for a useful proof visualization tool, its ability to support different viewers and the switching between them.

%

\section{Integration in \pt}
\label{sec:pt}

We have integrated the Sunburst view as an option accessible from the menu. Choosing this option loads the Sunburst view into a separate dialog window.
In addition to the proof, which is displayed on the left side, we display also an inference panel.
The information in this panel contains details about the inference: its type, its primary and auxiliary formulas, and quantifier instantiation information, if applicable.

In the remainder of this section, we will describe the new interface in terms of the conditions given in Section~\ref{sec:criteria}.
To emphasize what is written, we have inserted snapshots of views of actual proofs. The appendix contains
information about the names of each proof and of how to load it using our system.

\begin{description}
  \item[Displaying of large proofs.] The traditional Gentzen layout contains abundant and redundant white\linebreak spaces, not only due to its use edges, but also because it has to create extra horizontal spaces between premises of binary inferences.
    Therefore, proofs with many binary inferences, even if the formulas are hidden, are too wide to fit on the screen.
    An example of this can be seen in Figure~\ref{fig:tape_LK}.

\begin{figure}[hbt]
 \centering
 \includegraphics[width=0.8\textwidth]{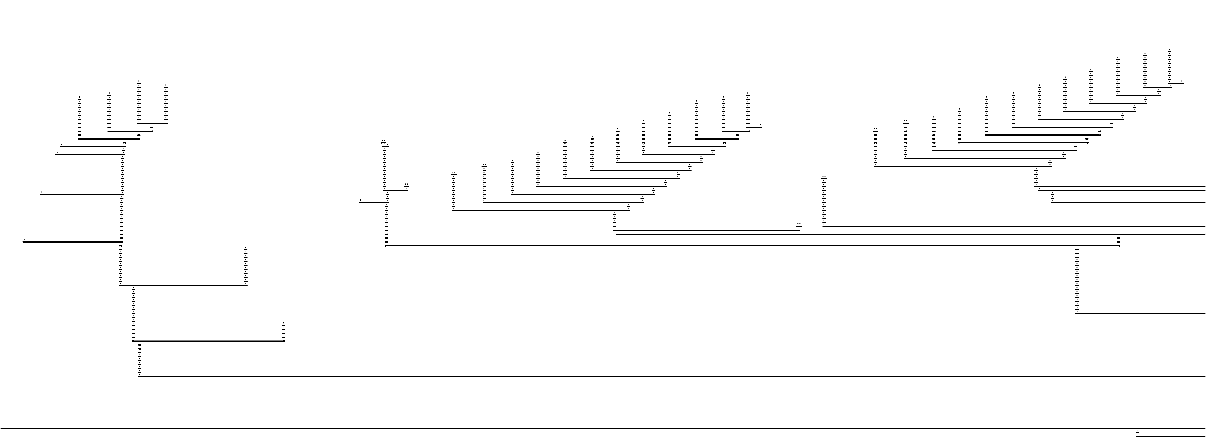}
 \caption{A small part of the Gentzen view of a proof with more than 2000 nodes.}
 \label{fig:tape_LK}
\end{figure}

  Projecting the sequent calculus proof to a circle allows roughly four times more space to render the inferences of a certain
  level\footnote{Let us assume the Gentzen proof to be an isosceles triangle with base length $w_1$, and height $d$.
    If we further assume the window is maximized, we can estimate the ratio $w_1:d=16:10$.
    Then the height $d$ is also the radius of the Sunburst tree giving it circumference $w_2 = 2 d \pi$.
    The ratio of the two lengths is then $w_1 : w_2 = \frac{20\pi}{16} \approx 4$.  }.
  Since the formulas are hidden, an inference is an easily clickable section of the disc, covering the whole area below its parents.
  Also, hovering over an inference with the mouse cursor triggers a darkening of its bounds.
  This is particularly helpful when tracing a formula throughout a proof, as one can then easily identify branching without the need to zoom in.

 As an example, Figure~\ref{fig:tape} shows a proof with more than 2000 inferences together with a zoomed in subproof.
 Comparing the two figures shows that even if we hide all inference information in the Gentzen view,
 we will still not be able to fit this proof on the screen.
 In contrast, the Sunburst view allows us to identify the main parts which constitute the proof:
 the top and bottom side have the same shape, for they contain the same reasoning structure on different terms.
 Only a small proof, which gives rise to a case distinction, is situated on the left hand side.
 Just by hovering over or selecting the cut-formulas (colored green), the user can identify the three parts of the proof.
 The right hand side of Figure~\ref{fig:tape} shows a zoom into one of the proof instances just mentioned.

\begin{figure}[hbt]
 \centering
 \includegraphics[width=0.48\textwidth]{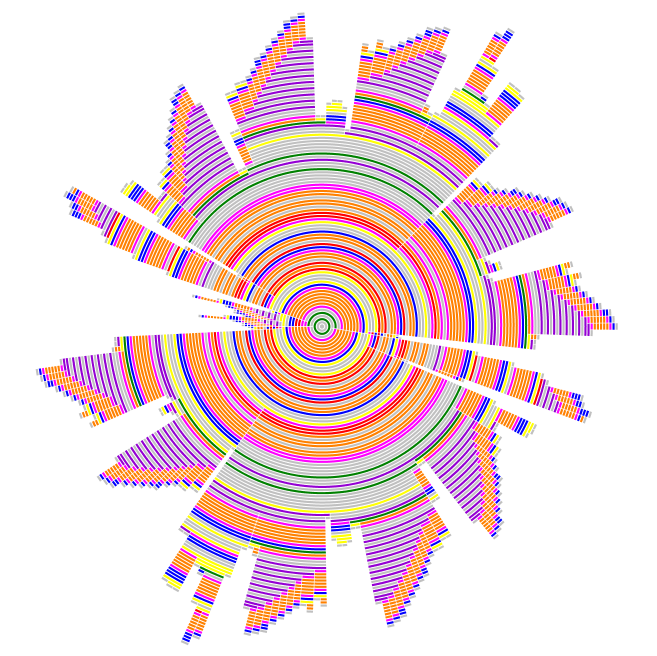}
 \includegraphics[width=0.48\textwidth]{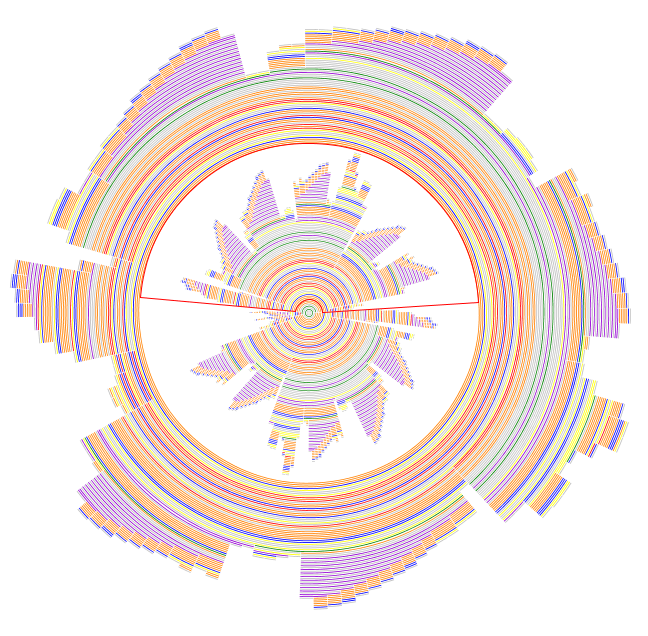}
 \caption{Sunburst view of a large proof in full view (left) and zoomed in (right).}
 \label{fig:tape}
\end{figure}

\item[Distinguishing between different kinds of rules.] This is easily achieved in Sunburst view by coloring the inference depending on the rule type.
  In order to obtain the highest contrast, we assigned the colors of the rainbow (see Figures \ref{fig:tape} and \ref{fig:zooming}) to groups of rules according to Table \ref{tab:colors}.
  \begin{table}[!ht]
    \centering
    \begin{tabular}[h]{|ll||ll||ll|}
      \hline
      Cut & green & Unary Logical Rule & orange & Strong Quantifier Rule & red \\
      Structural Rule & gray & Binary Logical Rule & yellow &  Weak Quantifier Rule & blue \\
      Axiom & gray & Equational Rule & violet & anything else & magenta\\
      \hline
    \end{tabular}
    \caption{Rules coloring schemes.}
    \label{tab:colors}
  \end{table}
  The relative size of subtrees to each other is adjustable by defining a so called weight. At the moment, the weight of a subtree is just the number of inferences. Depending on the application, one can imagine metrics which prioritize specific rules or even specific inferences.

\begin{figure}[hbt]
  \centering
  \includegraphics[width=0.45\textwidth]{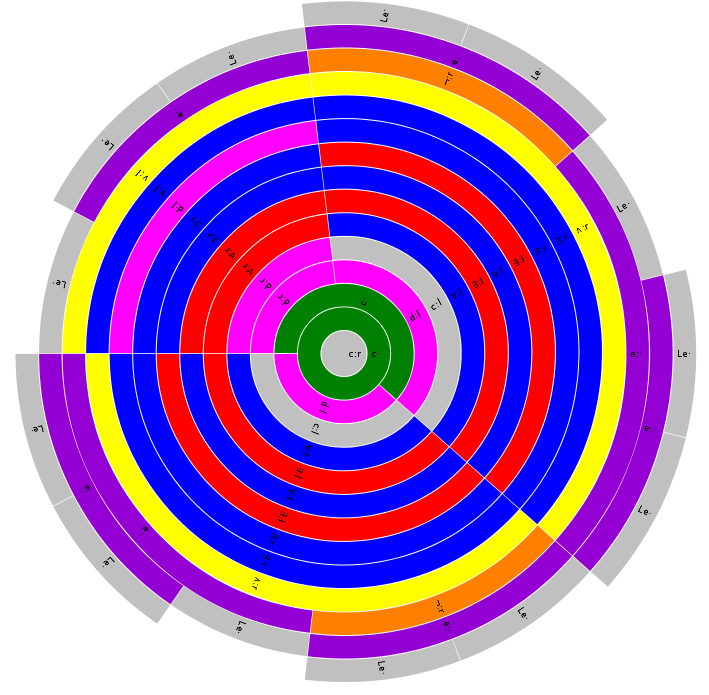}
  \includegraphics[width=0.45\textwidth]{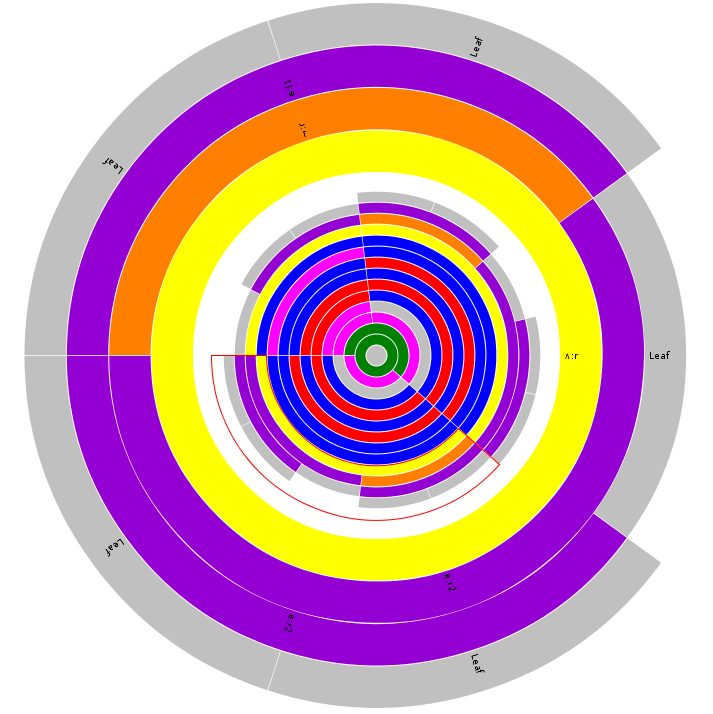}
  \caption{A combinatory proof (left) with a zoom into one of its subproofs (right).}
  \label{fig:zooming}
\end{figure}

  \item[Easy navigation.] Navigation is very easy in Sunburst, as can be seen in Figure~\ref{fig:zooming}.
    A single click on a subproof shrinks the whole proof to half of its size while displaying the subproof
    on its outer ring. Navigating backward can be done using the nested original proof.
    There are also keyboard shortcuts (Ctrl + arrow keys) available in the Sunburst view in order to help users navigate inside the proof.
    The up key moves selection to the child of a unary inference; the left and right keys select their respective premises of a binary inference and the down key selects the parent of any currently selected inference.

    One drawback of this form of navigation is that zooming into a subproof distorts its shape, affecting the user ability to understand the structure of the proof.
    Zooming into a subproof distorts all inferences in the same way.
    We therefore believe that this has only a minor affect on understanding the proof structure, since a human user can easily compensate for this
    fixed distortion.

  The Gentzen layout and the Sunburst view are tightly integrated to make better use of their respective strengths.
  When one navigates using the keyboard inside the proof in Sunburst view, the Gentzen viewer scrolls to the end-sequent of the selected node and highlights
  it as shown in Figure~\ref{fig:comparison}. There is also a ``Show node in LK view'' context menu available in the Sunburst view which allows to scroll to the chosen node in the Gentzen view.

\begin{figure}[hbt]
 \centering
 \includegraphics[scale=.35]{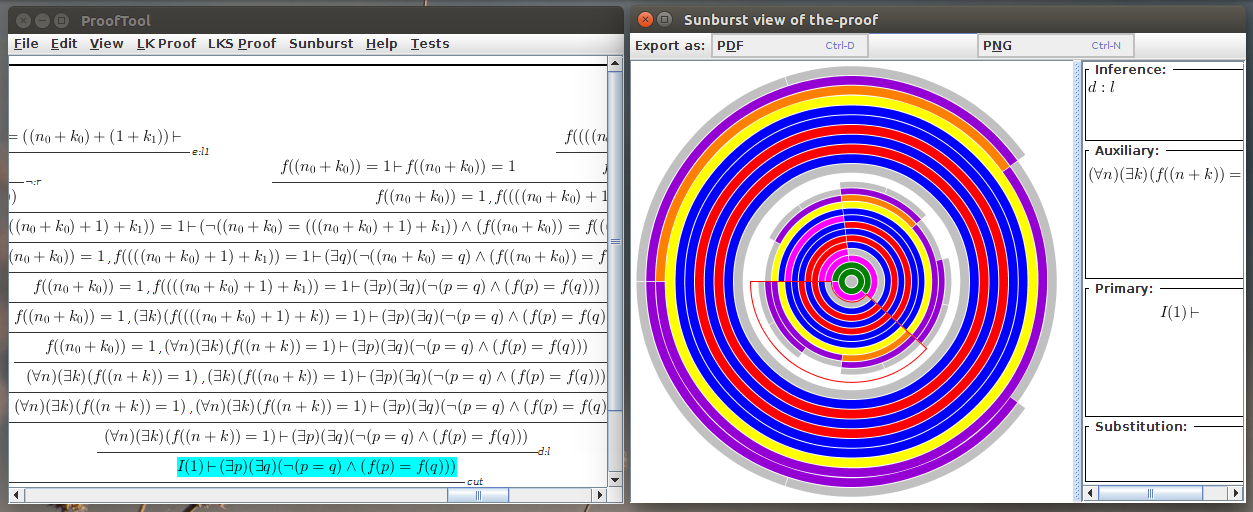} 
 \caption{Synchronizing the two views.}
 \label{fig:comparison}
\end{figure}

  \item[Formula ancestor information.] The sunburst view window is divided into two parts as shown in Figure~\ref{fig:comparison}.
    The first part shows the structure of the proof, while the second part gives additional information about the selected node.
    This includes the inference name, its auxiliary and principal formulas, and the substitution used, if any.
    But still, this information is not enough to see the ancestor relationship as well as it is possible in the Gentzen layout.
    The two views complement each other in this aspect as is illustrated in Figure~\ref{fig:ancestors}.

\begin{figure}[hbt]
 \centering
 \includegraphics[scale=.35]{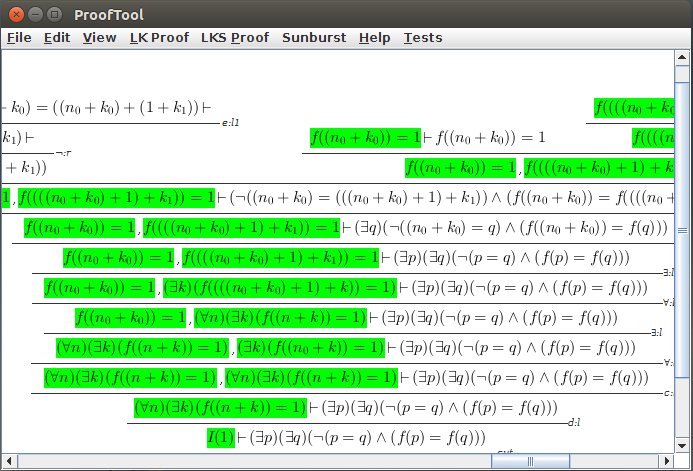} 
 \caption{The cut-formula ancestors marked green.}
 \label{fig:ancestors}
\end{figure}

  \item[Focus on different aspects of the proof.] In order to display different aspects of the same proof, one can take advantage of the possibility
    to customize the colors and width ratio of inferences in Sunburst. We would like to have a set of such pre-defined customizations which will
    emphasize different aspects, such as sub-proof and cut complexities, variable instantiations and specific rules and inferences.
    We plan to implement this feature in the near future.

  \item[Shape of proofs.]
    In some situations where the formula level is obscured, it is helpful to concentrate on the structure of the proof. In the following we describe two phenomenons we encountered.
    \begin{description}
    \item [Proof transformations often keep the structure intact.]
      Some proof transformations such as\linebreak elimination of definition rules and skolemization strongly change the proof on a formula level, but only slightly modify the structural layout.
      Nonetheless, locating an inference in the Gentzen layout with the find function of \pt\ becomes virtually impossible,
      since the subterms allowing a unique identification of a formula often have changed.
      The Sunburst view allows to use the proof structure to find the inference.
      For example, it is not always clear how a skolem term ends up in a weak quantifier inference,
      since the term might be carried over from a different part of the proof.
      In the Sunburst view, we can navigate to this inference and use both views for further investigation.

    \item [Understanding proof arguments.]
      In the process of formalizing a proof, one might not recognize all the possibilities where the proof can be generalized. In the Sunburst view, structural similarities are easier to spot and can then be checked whether a generalization is indeed possible.

      As an example, we can look at subsequent instances of a formalization of  Fürstenberg's proof of the infinity of primes~\cite{DBLP:journals/tcs/BaazHLRS08}. Here the schematic nature of the proof  was already taken into account during formalization, but now the induction argument becomes clearly visible (see Figure~\ref{fig:prime}).

      \begin{figure}[hbt]
        \centering
        \includegraphics[width=0.32\textwidth]{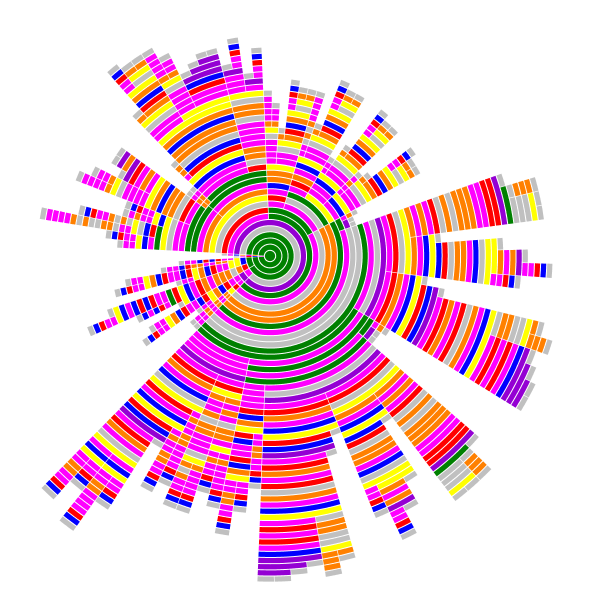}
        \includegraphics[width=0.32\textwidth]{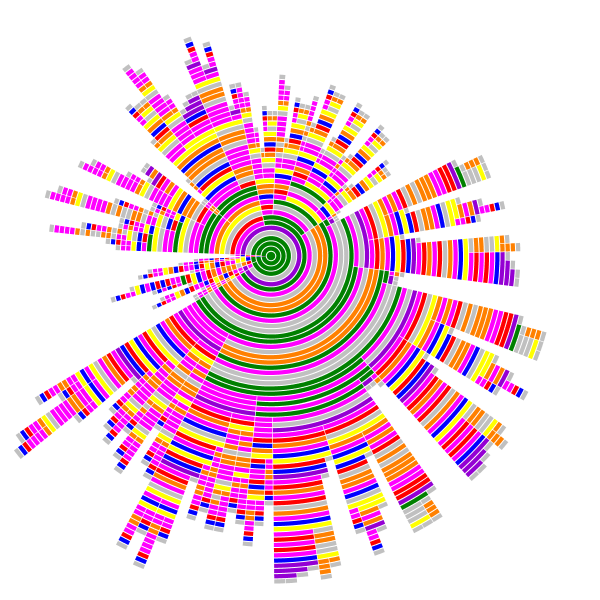}
        \includegraphics[width=0.32\textwidth]{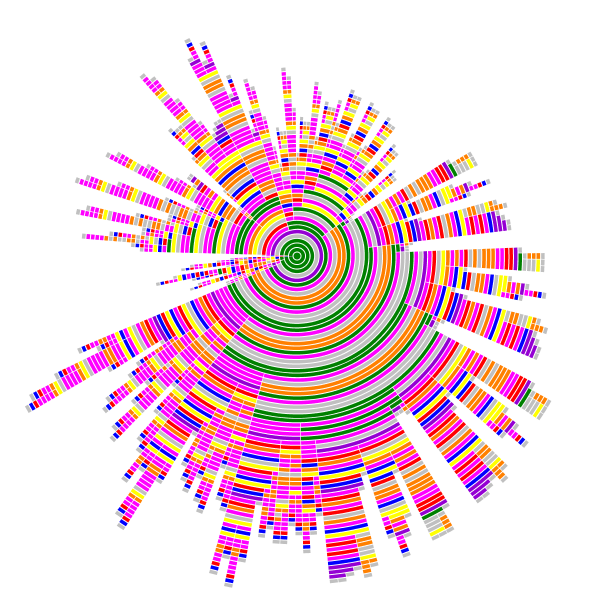}
        \caption{Instances 1, 2 and 3 of the formalization of Fürstenberg's proof of the infinity of primes.}
        \label{fig:prime}
      \end{figure}
    \end{description}

\end{description}



\section{Implementation}\label{sec:imp}

The GAPT framework is implemented in the programming language Scala~\cite{Oderski2010}. \pt makes heavy use of Scala's Swing wrapper library. Details about \pt and how it displays formulas, sequents and proofs can be found in~\cite{UITP}. In this section we concentrate on the integration of the Sunburst view in \pt.

Our implementation is based on \textsc{TreeViz}, an open source library\footnote{Visualization of Large Tree Structures, \url{http://www.randelshofer.ch/treeviz/}} written in Java. Since\linebreak \pt, and GAPT in general, are implemented in Scala, we have several wrappers around the classes from the \textsc{TreeViz} library. They also add functionality in the form of events which expose a newly selected node.

\begin{figure}[h]
 \centering
 \includegraphics[width=\textwidth]{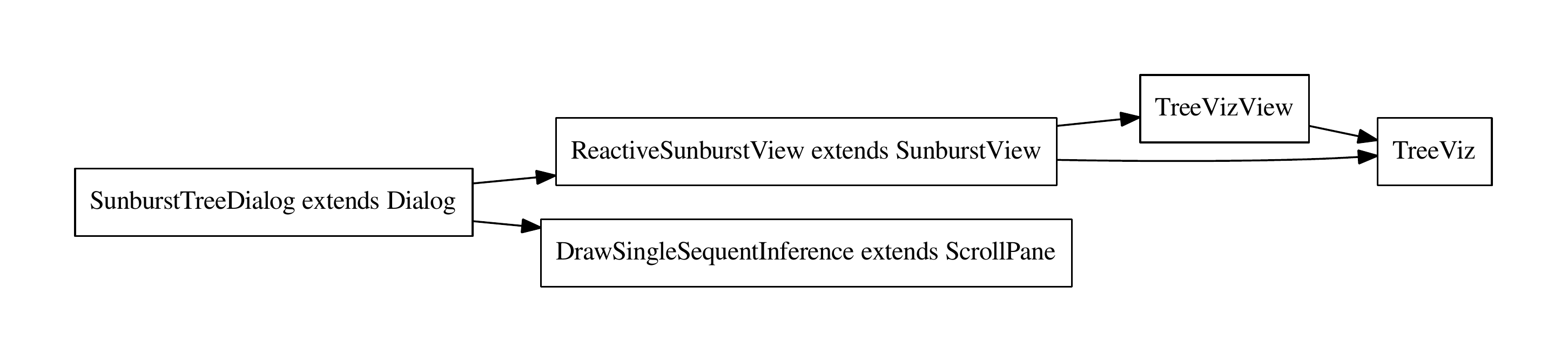}
 \vskip-2em
 \caption{Architecture of \textsc{SunburstTreeDialog}}
 \label{fig:architecture}
\end{figure}

We created a \textsc{SunburstTreeDialog} that displays the structural information as a Sunburst tree as well as the information about the selected inference. The full architecture of \textsc{SunburstTreeDialog} is shown in Figure~\ref{fig:architecture}

\textsc{SunburstTreeDialog} consists of a \textsc{SplitPane}, containing our Sunburst wrapper \textsc{Reactive\-SunburstView} on the left (upper) side. The right (lower) side contains the inference viewer called \textsc{DrawSingleSequentInference}. The orientation of the split pane is detected on run-time, depending on the dialog window size, allocating maximal space to the Sunburst view of the proof. The user can change this by either resizing the window or moving the delimiter of the pane.


\textsc{DrawSingleSequentInference} extends a \textsc{ScrollPane} and uses \textsc{BoxLayout} to display information about the selected inference, such as the inference name, auxiliary and principal formulas, and in cases of weak quantifier rules, it displays the substitution as well.
In order to fit formulas  best on the screen when the main window aligns objects horizontally, the \textsc{DrawSingleSequentInference} layout is changing the orientation from vertical to horizontal and vice versa, depending on the size of the auxiliary and principal formulas. This means that shorter formulas are aligned side-by-side and longer ones on top of each other.

\section{Conclusion and Future Work}\label{sec:conc}

In this paper we have explained the issues that standard tree viewers have when faced with large proofs.
We have then identified various criteria for a suitable tree visualization and analyzed the available algorithms with respect to them.
Our results show that Sunburst Trees seem to be the most adequate structural layout for viewing sequent calculus proofs.
The global structure can be better seen than in standard layouts, which makes large proofs readable.
We found identifying inferences, navigation, and tracing derivations superior to the Gentzen layout.
At the same time, formula or context intensive tasks such as identifying the ancestor relationship are better left to the latter.
The integration of the Sunburst viewer alongside the Gentzen viewer in \pt demonstrate how well these two complementary layouts interact with each other.

Some improvements are still of interest to us.
Foremost, multiple Sunburst trees can be represented by a forest structure. We plan to take advantage of this in two ways.
First, larger proofs usually consist of several subproofs solving partial problems.
In other words, they can be represented as a forest with links~\cite{dunchev13} to their subproofs.
This division is often explicitly contained in the proof input
language\footnote{The proof languages hlk, shlk and llk defined in the context of GAPT can be seen as examples for this.},
but the proof object itself usually does not carry on this information.
If all proof transformations are adjusted to carry on the link structure, the result can be divided into a set of proofs, making the meta-structure of the proof visible.
Moreover, it is also possible to display DAGs by converting them into forests. This would enable the viewer to display resolution refutations.
The high reuse of clauses in a refutation might fill the forest with many tiny trees, but this is open to experimentation.

A practical improvement is the addition of viewing profiles. By setting different color schemes and inference width ratios for each profile,
we can customize the viewer to better display different aspects of proofs, like subproof complexity and instantiations.

The last two planned improvements are on the level of formulas. First, we might increase the readability of large formulas by replacing them
with new symbols. In addition, we would like to improve the search facilities in \pt. Right now, searching for a specific formula in the
Gentzen view mark all occurrences of the formula. We plan to add a similar facility to the Sunburst view.
One idea is to put the search results into a list in a new window, thus allowing the user to browse through the search results and jump to the right inference.

\section*{Acknowledgements}

We would like to thank the anonymous reviewers for their helpful comments and suggestions.\\ \pt also benefited substantially from the participant's feedback which came up during discussions in the 11th UITP Workshop in Vienna.

\bibliographystyle{eptcs}
\bibliography{refs}

\begin{thebibliography}{10}
\providecommand{\bibitemdeclare}[2]{}
\providecommand{\surnamestart}{}
\providecommand{\surnameend}{}
\providecommand{\urlprefix}{Available at }
\providecommand{\url}[1]{\texttt{#1}}
\providecommand{\href}[2]{\texttt{#2}}
\providecommand{\urlalt}[2]{\href{#1}{#2}}
\providecommand{\doi}[1]{doi:\urlalt{http://dx.doi.org/#1}{#1}}
\providecommand{\bibinfo}[2]{#2}

\bibitemdeclare{inproceedings}{aspinall07}
\bibitem{aspinall07}
\bibinfo{author}{David \surnamestart Aspinall\surnameend},
  \bibinfo{author}{Christoph \surnamestart L\"{u}th\surnameend} \&
  \bibinfo{author}{Daniel \surnamestart Winterstein\surnameend}
  (\bibinfo{year}{2007}): \emph{\bibinfo{title}{A Framework for Interactive
  Proof}}.
\newblock In: {\sl \bibinfo{booktitle}{Proceedings of the 14th Symposium on
  Towards Mechanized Mathematical Assistants: 6th International Conference}},
  \bibinfo{publisher}{Springer-Verlag}, \bibinfo{address}{Berlin, Heidelberg},
  pp. \bibinfo{pages}{161--175}, \doi{10.1007/978-3-540-73086-6\_15}.

\bibitemdeclare{inproceedings}{Baaz2005}
\bibitem{Baaz2005}
\bibinfo{author}{Matthias \surnamestart Baaz\surnameend},
  \bibinfo{author}{Stefan \surnamestart Hetzl\surnameend},
  \bibinfo{author}{Alexander \surnamestart Leitsch\surnameend},
  \bibinfo{author}{Clemens \surnamestart Richter\surnameend} \&
  \bibinfo{author}{Hendrik \surnamestart Spohr\surnameend}
  (\bibinfo{year}{2005}): \emph{\bibinfo{title}{{Cut-Elimination: Experiments
  with CERES}}}.
\newblock In \bibinfo{editor}{Franz \surnamestart Baader\surnameend} \&
  \bibinfo{editor}{Andrei \surnamestart Voronkov\surnameend}, editors: {\sl
  \bibinfo{booktitle}{Logic for Programming, Artificial Intelligence, and
  Reasoning (LPAR) 2004}}, {\sl \bibinfo{series}{Lecture Notes in Computer
  Science}} \bibinfo{volume}{3452}, \bibinfo{publisher}{Springer}, pp.
  \bibinfo{pages}{481--495}, \doi{10.1007/978-3-540-32275-7\_32}.

\bibitemdeclare{article}{DBLP:journals/tcs/BaazHLRS08}
\bibitem{DBLP:journals/tcs/BaazHLRS08}
\bibinfo{author}{Matthias \surnamestart Baaz\surnameend},
  \bibinfo{author}{Stefan \surnamestart Hetzl\surnameend},
  \bibinfo{author}{Alexander \surnamestart Leitsch\surnameend},
  \bibinfo{author}{Clemens \surnamestart Richter\surnameend} \&
  \bibinfo{author}{Hendrik \surnamestart Spohr\surnameend}
  (\bibinfo{year}{2008}): \emph{\bibinfo{title}{CERES: An analysis of
  F{\"u}rstenberg's proof of the infinity of primes}}.
\newblock {\sl \bibinfo{journal}{Theor. Comput. Sci.}}
  \bibinfo{volume}{403}(\bibinfo{number}{2-3}), pp. \bibinfo{pages}{160--175},
  \doi{10.1016/j.tcs.2008.02.043}.

\bibitemdeclare{article}{battista1994algorithms}
\bibitem{battista1994algorithms}
\bibinfo{author}{Giuseppe~Di \surnamestart Battista\surnameend},
  \bibinfo{author}{Peter \surnamestart Eades\surnameend},
  \bibinfo{author}{Roberto \surnamestart Tamassia\surnameend} \&
  \bibinfo{author}{Ioannis~G \surnamestart Tollis\surnameend}
  (\bibinfo{year}{1994}): \emph{\bibinfo{title}{Algorithms for drawing graphs:
  an annotated bibliography}}.
\newblock {\sl \bibinfo{journal}{Computational Geometry}}
  \bibinfo{volume}{4}(\bibinfo{number}{5}), pp. \bibinfo{pages}{235--282},
  \doi{10.1016/0925-7721(94)00014-X}.

\bibitemdeclare{article}{cain2010}
\bibitem{cain2010}
\bibinfo{author}{AlanJ. \surnamestart Cain\surnameend} (\bibinfo{year}{2010}):
  \emph{\bibinfo{title}{Deus ex Machina and the Aesthetics of Proof}}.
\newblock {\sl \bibinfo{journal}{The Mathematical Intelligencer}}
  \bibinfo{volume}{32}(\bibinfo{number}{3}), pp. \bibinfo{pages}{7--11},
  \doi{10.1007/s00283-010-9141-z}.

\bibitemdeclare{inproceedings}{PxTP}
\bibitem{PxTP}
\bibinfo{author}{Cvetan \surnamestart Dunchev\surnameend},
  \bibinfo{author}{Alexander \surnamestart Leitsch\surnameend},
  \bibinfo{author}{Tomer \surnamestart Libal\surnameend},
  \bibinfo{author}{Martin \surnamestart Riener\surnameend},
  \bibinfo{author}{Mikheil \surnamestart Rukhaia\surnameend},
  \bibinfo{author}{Daniel \surnamestart Weller\surnameend} \&
  \bibinfo{author}{Bruno \surnamestart Woltzenlogel-Paleo\surnameend}
  (\bibinfo{year}{2012}): \emph{\bibinfo{title}{{System Feature Description:
  Importing Refutations into the GAPT Framework}}}.
\newblock In \bibinfo{editor}{David \surnamestart Pichardie\surnameend} \&
  \bibinfo{editor}{Tjark \surnamestart Weber\surnameend}, editors: {\sl
  \bibinfo{booktitle}{Second International Workshop on Proof Exchange for
  Theorem Proving (PxTP 2012)}}, {\sl \bibinfo{series}{CEUR Workshop
  Proceedings}} \bibinfo{volume}{878}, pp. \bibinfo{pages}{51--57}.

\bibitemdeclare{inproceedings}{UITP}
\bibitem{UITP}
\bibinfo{author}{Cvetan \surnamestart Dunchev\surnameend},
  \bibinfo{author}{Alexander \surnamestart Leitsch\surnameend},
  \bibinfo{author}{Tomer \surnamestart Libal\surnameend},
  \bibinfo{author}{Martin \surnamestart Riener\surnameend},
  \bibinfo{author}{Mikheil \surnamestart Rukhaia\surnameend},
  \bibinfo{author}{Daniel \surnamestart Weller\surnameend} \&
  \bibinfo{author}{Bruno \surnamestart Woltzenlogel-Paleo\surnameend}
  (\bibinfo{year}{2013}): \emph{\bibinfo{title}{{ProofTool: a GUI for the GAPT
  Framework}}}.
\newblock In \bibinfo{editor}{Cezary \surnamestart Kaliszyk\surnameend} \&
  \bibinfo{editor}{Christoph \surnamestart L\"uth\surnameend}, editors: {\sl
  \bibinfo{booktitle}{Proceedings 10th International Workshop On User
  Interfaces for Theorem Provers (UITP 2012)}}, {\sl
  \bibinfo{series}{Electronic Proceedings in Theoretical Computer Science}}
  \bibinfo{volume}{118}, pp. \bibinfo{pages}{1--14}, \doi{10.4204/EPTCS.118.1}.

\bibitemdeclare{article}{dunchev13}
\bibitem{dunchev13}
\bibinfo{author}{Cvetan \surnamestart Dunchev\surnameend},
  \bibinfo{author}{Alexander \surnamestart Leitsch\surnameend},
  \bibinfo{author}{Mikheil \surnamestart Rukhaia\surnameend} \&
  \bibinfo{author}{Daniel \surnamestart Weller\surnameend}
  (\bibinfo{year}{2013}): \emph{\bibinfo{title}{CERES for First-Order
  Schemata}}.
\newblock {\sl \bibinfo{journal}{CoRR}} \bibinfo{volume}{abs/1303.4257}.
\newblock \urlprefix\url{http://arxiv.org/abs/1303.4257}.

\bibitemdeclare{book}{hardy1992mathematician}
\bibitem{hardy1992mathematician}
\bibinfo{author}{Godfrey~Harold \surnamestart Hardy\surnameend}
  (\bibinfo{year}{1940}): \emph{\bibinfo{title}{A mathematician's apology}}.
\newblock \bibinfo{publisher}{Cambridge University Press}.

\bibitemdeclare{inproceedings}{HLRR13}
\bibitem{HLRR13}
\bibinfo{author}{Stefan \surnamestart Hetzl\surnameend}, \bibinfo{author}{Tomer
  \surnamestart Libal\surnameend}, \bibinfo{author}{Martin \surnamestart
  Riener\surnameend} \& \bibinfo{author}{Mikheil \surnamestart
  Rukhaia\surnameend} (\bibinfo{year}{2013}):
  \emph{\bibinfo{title}{{Understanding Resolution Proofs through Herbrand's
  Theorem}}}.
\newblock In \bibinfo{editor}{Didier \surnamestart Galmiche\surnameend} \&
  \bibinfo{editor}{Dominique \surnamestart Larchey-Wendling\surnameend},
  editors: {\sl \bibinfo{booktitle}{Automated Reasoning with Analytic Tableaux
  and Related Methods (Tableaux 2013)}}, {\sl \bibinfo{series}{Lecture Notes in
  Computer Science}} \bibinfo{volume}{8123}, pp. \bibinfo{pages}{157--171},
  \doi{10.1007/978-3-642-40537-2\_15}.

\bibitemdeclare{}{Hida2005}
\bibitem{Hida2005}
\bibinfo{author}{Yozo \surnamestart Hida\surnameend}, \bibinfo{author}{John~O.
  \surnamestart Lamping\surnameend} \& \bibinfo{author}{Ramana~B. \surnamestart
  Rao\surnameend} (\bibinfo{year}{2005}): \emph{\bibinfo{title}{Tree
  visualization system and method based upon a compressed half-plane model of
  hyperbolic geometry}}.
\newblock \urlprefix\url{http://www.freepatentsonline.com/6901555.html}.

\bibitemdeclare{inproceedings}{huang2007}
\bibitem{huang2007}
\bibinfo{author}{Mao~Lin \surnamestart Huang\surnameend},
  \bibinfo{author}{Quang~Vinh \surnamestart Nguyen\surnameend},
  \bibinfo{author}{Wei \surnamestart Lai\surnameend} \& \bibinfo{author}{Xiaodi
  \surnamestart Huang\surnameend} (\bibinfo{year}{2007}):
  \emph{\bibinfo{title}{Three-Dimensional {EncCon} Tree}}.
\newblock In \bibinfo{editor}{Ebad \surnamestart Banissi\surnameend},
  \bibinfo{editor}{Muhammad \surnamestart Sarfraz\surnameend} \&
  \bibinfo{editor}{Natasha \surnamestart Dejdumrong\surnameend}, editors: {\sl
  \bibinfo{booktitle}{CGIV'07: Proceedings of the Computer Graphics, Imaging
  and Visualisation}}, \bibinfo{publisher}{IEEE Computer Society}, pp.
  \bibinfo{pages}{429--433}, \doi{10.1109/CGIV.2007.82}.

\bibitemdeclare{techreport}{huet97}
\bibitem{huet97}
\bibinfo{author}{G{\'e}rard \surnamestart Huet\surnameend},
  \bibinfo{author}{Gilles \surnamestart Kahn\surnameend} \&
  \bibinfo{author}{Christine \surnamestart Paulin-Mohring\surnameend}
  (\bibinfo{year}{1997}): \emph{\bibinfo{title}{{The Coq Proof Assistant : A
  Tutorial : Version 6.1}}}.
\newblock \bibinfo{type}{Rapport de recherche} \bibinfo{number}{RT-0204},
  \bibinfo{institution}{INRIA}.
\newblock \bibinfo{note}{Projet COQ}.

\bibitemdeclare{article}{knuth1971}
\bibitem{knuth1971}
\bibinfo{author}{D.E. \surnamestart Knuth\surnameend} (\bibinfo{year}{1971}):
  \emph{\bibinfo{title}{Optimum binary search trees}}.
\newblock {\sl \bibinfo{journal}{Acta Informatica}}
  \bibinfo{volume}{1}(\bibinfo{number}{1}), pp. \bibinfo{pages}{14--25},
  \doi{10.1007/BF00264289}.

\bibitemdeclare{inproceedings}{Lamping1995}
\bibitem{Lamping1995}
\bibinfo{author}{John \surnamestart Lamping\surnameend},
  \bibinfo{author}{Ramana \surnamestart Rao\surnameend} \&
  \bibinfo{author}{Peter \surnamestart Pirolli\surnameend}
  (\bibinfo{year}{1995}): \emph{\bibinfo{title}{A focus+context technique based
  on hyperbolic geometry for visualizing large hierarchies}}.
\newblock In \bibinfo{editor}{Irvin~R. \surnamestart Katz\surnameend},
  \bibinfo{editor}{Robert \surnamestart Mack\surnameend}, \bibinfo{editor}{Linn
  \surnamestart Marks\surnameend}, \bibinfo{editor}{Mary~Beth \surnamestart
  Rosson\surnameend} \& \bibinfo{editor}{Jakob \surnamestart
  Nielsen\surnameend}, editors: {\sl \bibinfo{booktitle}{CHI'95: Proceedings of
  the SIGCHI Conference on Human Factors in Computing Systems}},
  \bibinfo{publisher}{ACM Press/Addison-Wesley Publishing Co.}, pp.
  \bibinfo{pages}{401--408}, \doi{10.1145/223904.223956}.

\bibitemdeclare{article}{Linsen2011}
\bibitem{Linsen2011}
\bibinfo{author}{Lars \surnamestart Linsen\surnameend} \&
  \bibinfo{author}{Sabine \surnamestart Behrendt\surnameend}
  (\bibinfo{year}{2011}): \emph{\bibinfo{title}{Linked Treemap: A {3D}
  Treemap-nodelink layout for visualizing hierarchical structures}}.
\newblock {\sl \bibinfo{journal}{Computational Statistics}}
  \bibinfo{volume}{26}(\bibinfo{number}{4}), pp. \bibinfo{pages}{679--697},
  \doi{10.1007/s00180-011-0272-2}.

\bibitemdeclare{article}{Meier1996}
\bibitem{Meier1996}
\bibinfo{author}{John \surnamestart Meier\surnameend} \&
  \bibinfo{author}{Clifford~A. \surnamestart Reiter\surnameend}
  (\bibinfo{year}{1996}): \emph{\bibinfo{title}{Fractal representations of
  {Cayley} graphs}}.
\newblock {\sl \bibinfo{journal}{Computers and Graphics}}
  \bibinfo{volume}{20}(\bibinfo{number}{1}), pp. \bibinfo{pages}{163--170},
  \doi{10.1016/0097-8493(95)00101-8}.

\bibitemdeclare{inproceedings}{munzner1997}
\bibitem{munzner1997}
\bibinfo{author}{Tamara \surnamestart Munzner\surnameend}
  (\bibinfo{year}{1997}): \emph{\bibinfo{title}{H3: laying out large directed
  graphs in {3D} hyperbolic space}}.
\newblock In \bibinfo{editor}{John \surnamestart Dill\surnameend} \&
  \bibinfo{editor}{Nahum~D. \surnamestart Gershon\surnameend}, editors: {\sl
  \bibinfo{booktitle}{InfoVis'97: Proceedings of the IEEE Symposium on
  Information Visualization}}, \bibinfo{publisher}{IEEE Computer Society}, pp.
  \bibinfo{pages}{2--10}, \doi{10.1109/INFVIS.1997.636718}.

\bibitemdeclare{book}{Oderski2010}
\bibitem{Oderski2010}
\bibinfo{author}{Martin \surnamestart Odersky\surnameend}, \bibinfo{author}{Lex
  \surnamestart Spoon\surnameend} \& \bibinfo{author}{Bill \surnamestart
  Venners\surnameend} (\bibinfo{year}{2010}): \emph{\bibinfo{title}{Programming
  in {S}cala: A Comprehensive Step-by-step Guide}}, \bibinfo{edition}{2nd}
  edition.
\newblock \bibinfo{publisher}{Artima, Inc.}

\bibitemdeclare{mastersthesis}{paterson2013}
\bibitem{paterson2013}
\bibinfo{author}{Grace~D \surnamestart Paterson\surnameend}
  (\bibinfo{year}{2013}): \emph{\bibinfo{title}{The Aesthetics of Mathematical
  Proofs}}.
\newblock Master's thesis.

\bibitemdeclare{article}{Rosindell2012}
\bibitem{Rosindell2012}
\bibinfo{author}{James \surnamestart Rosindell\surnameend} \&
  \bibinfo{author}{Luke~J. \surnamestart Harmon\surnameend}
  (\bibinfo{year}{2012}): \emph{\bibinfo{title}{{OneZoom}: A Fractal Explorer
  for the {Tree of Life}}}.
\newblock {\sl \bibinfo{journal}{PLoS Biology}}
  \bibinfo{volume}{10}(\bibinfo{number}{10}), p. \bibinfo{pages}{e1001406},
  \doi{10.1371/journal.pbio.1001406}.

\bibitemdeclare{inproceedings}{Rusu2007a}
\bibitem{Rusu2007a}
\bibinfo{author}{Adrian \surnamestart Rusu\surnameend} \&
  \bibinfo{author}{Confesor \surnamestart Santiago\surnameend}
  (\bibinfo{year}{2007}): \emph{\bibinfo{title}{A Practical Algorithm for
  Planar Straight-line Grid Drawings of General Trees with Linear Area and
  Arbitrary Aspect Ratio}}.
\newblock In \bibinfo{editor}{Ebad \surnamestart Banissi\surnameend},
  \bibinfo{editor}{Remo~Aslak \surnamestart Burkhard\surnameend},
  \bibinfo{editor}{Georges \surnamestart Grinstein\surnameend},
  \bibinfo{editor}{Urska \surnamestart Cvek\surnameend},
  \bibinfo{editor}{Marjan \surnamestart Trutschl\surnameend},
  \bibinfo{editor}{Liz \surnamestart Stuart\surnameend},
  \bibinfo{editor}{Theodor~G. \surnamestart Wyeld\surnameend},
  \bibinfo{editor}{Gennady \surnamestart Andrienko\surnameend},
  \bibinfo{editor}{Jason \surnamestart Dykes\surnameend},
  \bibinfo{editor}{Mikael \surnamestart Jern\surnameend},
  \bibinfo{editor}{Dennis \surnamestart Groth\surnameend} \&
  \bibinfo{editor}{Anna \surnamestart Ursyn\surnameend}, editors: {\sl
  \bibinfo{booktitle}{IV'07: Proceedings of the International Conference on
  Information Visualisation}}, \bibinfo{publisher}{IEEE Computer Society}, pp.
  \bibinfo{pages}{743--750}, \doi{10.1109/IV.2007.14}.

\bibitemdeclare{article}{schulz2011}
\bibitem{schulz2011}
\bibinfo{author}{H.~\surnamestart Schulz\surnameend} (\bibinfo{year}{2011}):
  \emph{\bibinfo{title}{Treevis.net: A Tree Visualization Reference}}.
\newblock {\sl \bibinfo{journal}{Computer Graphics and Applications, IEEE}}
  \bibinfo{volume}{31}(\bibinfo{number}{6}), pp. \bibinfo{pages}{11--15},
  \doi{10.1109/MCG.2011.103}.

\bibitemdeclare{article}{Shneiderman91}
\bibitem{Shneiderman91}
\bibinfo{author}{Ben \surnamestart Shneiderman\surnameend}
  (\bibinfo{year}{1991}): \emph{\bibinfo{title}{Tree visualization with
  Tree-maps: A 2-d space-filling approach}}.
\newblock {\sl \bibinfo{journal}{ACM Transactions on Graphics}}
  \bibinfo{volume}{11}, pp. \bibinfo{pages}{92--99},
  \doi{10.1145/102377.115768}.

\bibitemdeclare{inproceedings}{W1}
\bibitem{W1}
\bibinfo{author}{J{\"o}rg \surnamestart Siekmann\surnameend},
  \bibinfo{author}{Stephan \surnamestart Hess\surnameend},
  \bibinfo{author}{Christoph \surnamestart Benzm{\"u}ller\surnameend},
  \bibinfo{author}{Lassaad \surnamestart Cheikhrouhou\surnameend},
  \bibinfo{author}{Detlef \surnamestart Fehrer\surnameend},
  \bibinfo{author}{Armin \surnamestart Fiedler\surnameend},
  \bibinfo{author}{Helmut \surnamestart Horacek\surnameend},
  \bibinfo{author}{Michael \surnamestart Kohlhase\surnameend},
  \bibinfo{author}{Karsten \surnamestart Konrad\surnameend},
  \bibinfo{author}{Andreas \surnamestart Meier\surnameend},
  \bibinfo{author}{Erica \surnamestart Melis\surnameend} \&
  \bibinfo{author}{Volker \surnamestart Sorge\surnameend}
  (\bibinfo{year}{1998}): \emph{\bibinfo{title}{A Distributed Graphical User
  Interface for the Interactive Proof System}}.
\newblock In: {\sl \bibinfo{booktitle}{Proceedings of the International
  Workshop "User Interfaces for Theorem Provers 1998 (UITP'98)}},
  \bibinfo{address}{Eindhoven, Netherlands}, pp. \bibinfo{pages}{130--138}.
\newblock \urlprefix\url{http://christoph-benzmueller.de/papers/W1.pdf}.

\bibitemdeclare{inproceedings}{Stasko2000a}
\bibitem{Stasko2000a}
\bibinfo{author}{John \surnamestart Stasko\surnameend} \&
  \bibinfo{author}{Eugene \surnamestart Zhang\surnameend}
  (\bibinfo{year}{2000}): \emph{\bibinfo{title}{Focus+Context Display and
  Navigation Techniques for Enhancing Radial, Space-Filling Hierarchy
  Visualizations}}.
\newblock In \bibinfo{editor}{Jock~D. \surnamestart Mackinlay\surnameend},
  \bibinfo{editor}{Steven~F. \surnamestart Roth\surnameend} \&
  \bibinfo{editor}{Daniel~A. \surnamestart Keim\surnameend}, editors: {\sl
  \bibinfo{booktitle}{InfoVis'00: Proceedings of the IEEE Symposium on
  Information Visualization}}, \bibinfo{publisher}{IEEE Computer Society}, pp.
  \bibinfo{pages}{57--65}, \doi{10.1109/INFVIS.2000.885091}.

\bibitemdeclare{article}{SS98}
\bibitem{SS98}
\bibinfo{author}{G.~\surnamestart Sutcliffe\surnameend} (\bibinfo{year}{2009}):
  \emph{\bibinfo{title}{{The TPTP Problem Library and Associated
  Infrastructure: The FOF and CNF Parts, v3.5.0}}}.
\newblock {\sl \bibinfo{journal}{Journal of Automated Reasoning}}
  \bibinfo{volume}{43}(\bibinfo{number}{4}), pp. \bibinfo{pages}{337--362},
  \doi{10.1007/s10817-009-9143-8}.

\bibitemdeclare{article}{Trac2007109}
\bibitem{Trac2007109}
\bibinfo{author}{Steven \surnamestart Trac\surnameend}, \bibinfo{author}{Yury
  \surnamestart Puzis\surnameend} \& \bibinfo{author}{Geoff \surnamestart
  Sutcliffe\surnameend} (\bibinfo{year}{2007}): \emph{\bibinfo{title}{An
  Interactive Derivation Viewer}}.
\newblock {\sl \bibinfo{journal}{Electronic Notes in Theoretical Computer
  Science}} \bibinfo{volume}{174}(\bibinfo{number}{2}), pp. \bibinfo{pages}{109
  -- 123}, \doi{10.1016/j.entcs.2006.09.025}.
\newblock
  \urlprefix\url{http://www.sciencedirect.com/science/article/pii/S15710661070%
01739}.
\newblock \bibinfo{note}{Proceedings of the 7th Workshop on User Interfaces for
  Theorem Provers (UITP 2006)}.

\bibitemdeclare{article}{Wetherell1979}
\bibitem{Wetherell1979}
\bibinfo{author}{Charles \surnamestart Wetherell\surnameend} \&
  \bibinfo{author}{Alfred \surnamestart Shannon\surnameend}
  (\bibinfo{year}{1979}): \emph{\bibinfo{title}{Tidy Drawings of Trees}}.
\newblock {\sl \bibinfo{journal}{IEEE Transactions on Software Engineering}}
  \bibinfo{volume}{SE-5}(\bibinfo{number}{5}), pp. \bibinfo{pages}{514--520},
  \doi{10.1109/TSE.1979.234212}.

\bibitemdeclare{inproceedings}{RISC4536}
\bibitem{RISC4536}
\bibinfo{author}{W.~\surnamestart Windsteiger\surnameend}
  (\bibinfo{year}{2012}): \emph{\bibinfo{title}{{Theorema 2.0: A Graphical User
  Interface for a Mathematical Assistant System}}}.
\newblock In \bibinfo{editor}{Cezary \surnamestart Kaliszyk\surnameend} \&
  \bibinfo{editor}{Christoph \surnamestart Lueth\surnameend}, editors: {\sl
  \bibinfo{booktitle}{{Proceedings 10th International Workshop On User
  Interfaces for Theorem Provers, Bremen, Germany, July 11th 2012}}}, {\sl
  \bibinfo{series}{Electronic Proceedings in Theoretical Computer Science}}
  \bibinfo{volume}{118}, \bibinfo{publisher}{Open Publishing Association}, pp.
  \bibinfo{pages}{72--82}, \doi{10.4204/EPTCS.118.5}.
\newblock \urlprefix\url{http://arxiv.org/abs/1307.1945v1}.

\bibitemdeclare{inproceedings}{Yang2002}
\bibitem{Yang2002}
\bibinfo{author}{Jing \surnamestart Yang\surnameend},
  \bibinfo{author}{Matthew~O. \surnamestart Ward\surnameend} \&
  \bibinfo{author}{Elke~A. \surnamestart Rundensteiner\surnameend}
  (\bibinfo{year}{2002}): \emph{\bibinfo{title}{{InterRing}: An Interactive
  Tool for Visually Navigating and Manipulating Hierarchical Structures}}.
\newblock In \bibinfo{editor}{Pak~Chung \surnamestart Wong\surnameend} \&
  \bibinfo{editor}{Keith \surnamestart Andrews\surnameend}, editors: {\sl
  \bibinfo{booktitle}{InfoVis'02: Proceedings of the IEEE Symposium on
  Information Visualization}}, \bibinfo{publisher}{IEEE Computer Society}, pp.
  \bibinfo{pages}{77--84}, \doi{10.1109/INFVIS.2002.1173151}.

\bibitemdeclare{inproceedings}{Youn1988}
\bibitem{Youn1988}
\bibinfo{author}{Hee~Yong \surnamestart Youn\surnameend} \&
  \bibinfo{author}{Adit~D. \surnamestart Singh\surnameend}
  (\bibinfo{year}{1988}): \emph{\bibinfo{title}{Near optimal embedding of
  binary tree architecture in {VLSI}}}.
\newblock In: {\sl \bibinfo{booktitle}{ICDCS'88: Proceedings of the
  International Conference on Distributed Computing Systems}},
  \bibinfo{publisher}{IEEE Computer Society}, pp. \bibinfo{pages}{86--93},
  \doi{10.1109/DCS.1988.12503}.

\end{thebibliography}

\section*{Appendix A: How to use the Tool}\label{appendix}

All the \pt\ snapshots shown in this paper are taken using actual proofs.
In this appendix we will describe how to install the tool and display the proofs.

The GAPT framework provides two kinds of interfaces,
a commands based shell prompt and the graphical \pt.
The shell prompt is based on the Scala shell, so its usage is more comfortable if one needs to do inline programming apart from predefined functions.
To visualize generated objects it calls the \pt, which allows for both the presentation and manipulation of proofs and supports a subset
of the functionality of GAPT.

The two executables, bundled together with an examples directory, can be downloaded at \url{http://www.logic.at/gapt/downloads/gapt-1.9.zip}.
The examples directory contains, among other things, also the proofs used in this paper.

In this paper we used different versions of two proofs. The first proof, called the Tape proof,
proves that if there is an infinite tape filled-in with two symbols, then there are at least two cells with the same symbol~\cite{Baaz2005}.
The second proof is the formalization of F\"urstenberg’s proof of the infinity of primes~\cite{DBLP:journals/tcs/BaazHLRS08}.

In Figures~\ref{fig:tape_LK}~and~\ref{fig:tape} a higher-order version of the Tape proof is shown.
In order to load it from the shell prompt, first run the shell by executing \verb|./cli.sh| and then load the proof using the following commands:

\begin{verbatim}
scala> val p = loadLLK("./examples/hol-tape/ntape.llk")
scala> val elp = regularize(eliminateDefinitions(p, "TAPEPROOF"))._1
scala> val selp = lkTolksk(elp)
scala> PT.display("TAPEPROOF",selp)
\end{verbatim}

Figure~\ref{fig:tape_LK} is then obtained by changing the font size in the View menu to minimum and hiding all the formulas (from the Edit menu).
Figure~\ref{fig:tape} is obtained by calling ``Sunburst view'' form the Sunburst menu and selecting the inferences either by directly clicking on them
or by navigating to it via the keyboard.

The rest of the proofs can be loaded directly from \pt. To load the \pt\ directly, execute \verb|./gui.sh|.

The first-order Tape proof is shown in Figures~\ref{fig:zooming},~\ref{fig:comparison}~and~\ref{fig:ancestors}.
From the File menu, please open the \verb|./examples/tape/tape.xml.gz| file.
Open the Sunburst view and place the two windows side-by-side in order to see the connection between them.
Start navigating inside the Sunburst view from the keyboard and the other window will change its view as shown in Figure~\ref{fig:comparison}.
Call the Edit$>$Mark Cut-Ancestors menu item to get the effect shown in Figure~\ref{fig:ancestors}.

The F\"urstenberg’s proof is formalized as a proof schema~\cite{dunchev13} and instances for 1, 2 and 3 primes are extracted into separate files.
Using again the File menu, the three proofs can be loaded using the following files:

\begin{verbatim}
./examples/prime/ceres_xml/prime1-1.xml.gz
./examples/prime/ceres_xml/prime1-2.xml.gz
./examples/prime/ceres_xml/prime1-3.xml.gz
\end{verbatim}

and then calling the ``Sunburst view'' in order to get the views shown in Figure~\ref{fig:prime}.

\end{document}